\documentclass{iaus}
\usepackage{graphicx}

  \checkfont{eurm10}
  \iffontfound
    \IfFileExists{upmath.sty}
      {\typeout{^^JFound AMS Euler Roman fonts on the system,
                   using the 'upmath' package.^^J}%
       \usepackage{upmath}}
      {\typeout{^^JFound AMS Euler Roman fonts on the system, but you
                   dont seem to have the}%
       \typeout{'upmath' package installed. iaus.cls can take advantage
                 of these fonts,^^Jif you use 'upmath' package.^^J}%
      }
  \else
  \fi

  \checkfont{msam10}
  \iffontfound
    \IfFileExists{amssymb.sty}
      {\typeout{^^JFound AMS Symbol fonts on the system, using the
                'amssymb' package.^^J}%
       \usepackage{amssymb}%
         
         \let\geq=\geqslant
      }{}
  \fi

  \IfFileExists{amsbsy.sty}
    {\typeout{^^JFound the 'amsbsy' package on the system, using it.^^J}%
     \usepackage{amsbsy}}
    {\providecommand\boldsymbol[1]{\mbox{\boldmath $##1$}}}

\newsavebox{\astrutbox}
\sbox{\astrutbox}{\rule[-5pt]{0pt}{20pt}}

\title[The Interplay among Black Holes, Stars and ISM in Galactic 
       Nuclei]{Extended Coronal Emission Lines in Active Galactic Nuclei}

\author[A. Rodr\'{\i}guez-Ardila, A. Prieto, S. M. Viegas] %
{Alberto Rodr\'{\i}guez-Ardila$^1$, %
 Almudena Prieto$^2$,
\and Sueli M. Viegas$^3$}

\affiliation{$^1$ LNA/MCT, Rua dos Estados Unidos 154,
          Itajub\'a, MG, Brazil email: aardila@lna.br\\[\affilskip]
$^2$Max-Plank-Institut f\"ur Astronomie, Heidelberg, Germany \\[\affilskip]
$^3$IAG-Universidade de S\~ao Paulo, SP, Brazil}

\pubyear{2004}
\volume{222}
\pagerange{1--8}
\date{?? and in revised form ??}
\jname{The Interplay among Black Holes, Stars and ISM \\in Galactic Nuclei}
\editors{Th. Storchi Bergmann, L.C. Ho \& H.R. Schmitt, eds.}
\begin{document}

\maketitle

\begin{abstract}
VLT and NTT spectra are used to examine the nuclear and
extended coronal line emission in a sample of
well-known Seyfert 1 and 2 galaxies. The excellent
spatial resolution obtained with VLT allowed us to
map [Si\,{\sc vi}] 1.963$\mu$m and [Si\,{\sc vii}] 2.48$\mu$m 
on scales of up to 20 pc. Coronal line emission, extended to distances
of $\sim$10$^{2}$ pc, is detected in some of the
lines analyzed, particularly in [Fe\,{\sc x}] 6374\AA, 
[Fe\,{\sc xi}] 7891\AA, and [Si\,{\sc vii}] 2.48$\mu$m.
Most coronal lines are strongly
asymmetric towards the blue and broader than low-ionization
lines. This result is particularly important for Circinus,
where previous observations had failed at detecting larger
widths for high-ionization lines. Photoionization models are
used to investigate the physical conditions and
continuum luminosities necessary to produced the observed
coronal emission. We found that an ionization parameter $U>$ 0.10
is necessary to reproduce the observations, although the clouds
should be located at a distance $<$ 30\,pc.
\end{abstract}

\firstsection 

\section{Introduction}
Coronal lines (CL) are collisionally excited forbidden transitions 
within low-lying levels of highly ionized species (IP$>$100 eV).
They can be formed either by a hard UV continuum (Marconi et al. 1996; 
Ferguson et al. 1997), 
a very hot collisionally ionized plasma (Viegas-Aldrovandi \& Contini  
1989), or a combination of both processes (Contini et al. 1998).
Due to the high energies involved in their production, the
detection of CL is taken as 
an unambiguous signature of nuclear activity and can provide
clues on the UV to soft X-ray spectral energy
distribution of the active nucleus (Prieto et al. 2002). 

Observationally, CL are blueshifted relative to the systemic
velocity of the host galaxy ($\Delta$V $\sim$ 500-800 km/s) and
broader than low ionization lines (Penston et al. 
1984). This has led to the idea that CL are 
associated to outflows (Erkens et al. 1997) and formed in a separate 
region (termed as coronal line region, CLR) located between 
the classical narrow line region (NLR) and broad line region. 
However, both the physical conditions of the coronal gas
and the size of the emitting region are matters
of strong controversy (Erkens et al. 1997; Nazarova et al. 1999). 
Published results usually 
rely on measurements made on just a few CLs and 
information about the size of their emitting regions, on the same 
target, is very scarce.

With the above in mind, the goals of this work are:
{\it (i)} Observe a sample of well-known AGN in order to measure 
the size of the emitting region for different coronal lines;
{\it (ii)} model the observe emission to obtain clues on the physical 
conditions of the CLR; {\it (iii)} study the kinematics of the coronal gas 
and compare it to that of other NLR lines.  

\section{Main results} \label{res}

\begin{figure}
\centering
\includegraphics[width=6.0cm]{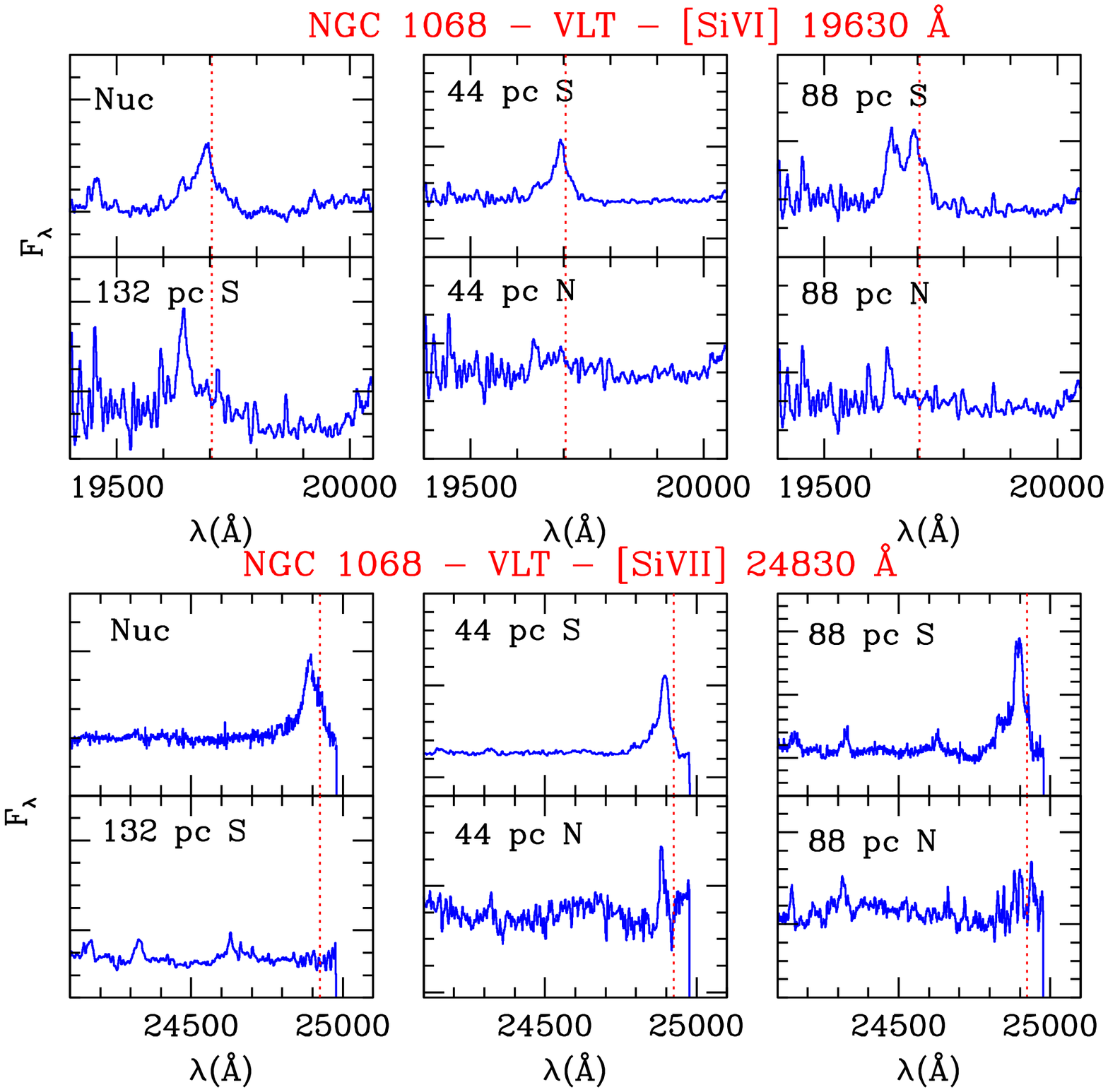}
\includegraphics[width=6.0cm]{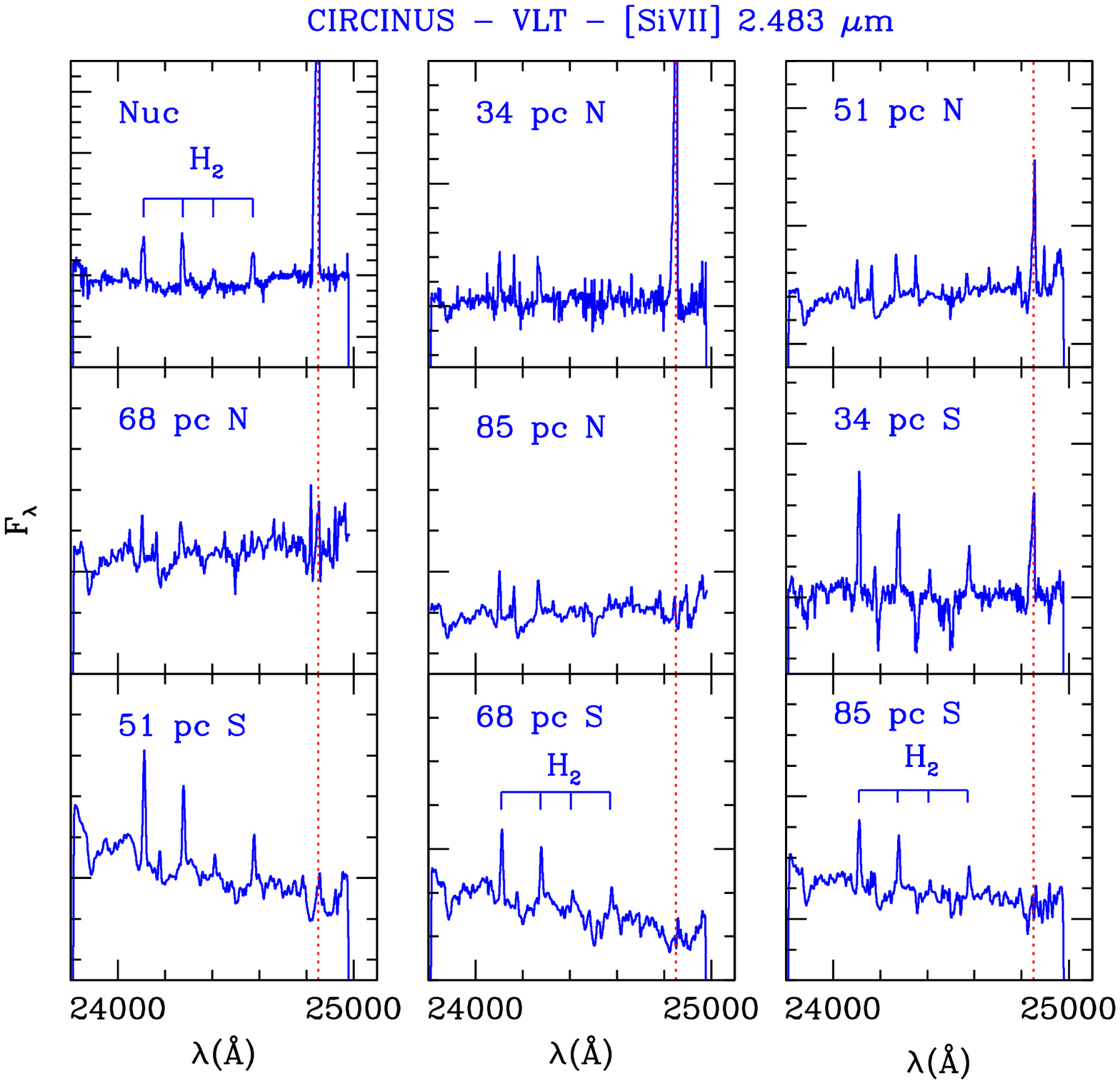}
  \caption{Observed VLT/ISAAC NIR spectra of NGC\,1068 (left) and 
Circinus (right). In each panel, the distance from the centre of the 
AGN to which the spectrum corresponds is indicated. The dotted line 
marks the centroid position of the CL. F$_{\lambda}$ is in arbitrary 
units.}\label{clr}
\end{figure}

A sub-sample of the observed data, collected using VLT/ISAAC 
at ESO for NGC\,1068 and Circinus, is presented in Figure~\ref{clr}.
The spatial scale of the detector (0.15''/pix), combined
with the good seeing during the observations ($\sim$0.6'') allowed
us to map gas emission at a spatial resolution down to 20\,pc. 
The slit in both configurations was oriented north-south.

The right panel of Figure~\ref{clr} shows that in Circinus the 
[Si\,{\sc vii}] line extends from the unresolved nucleus (R$<$17\,pc), 
up to $\sim$70\,pc to the north and 34\,pc to the south, 
implying a total size of $\sim$100\,pc for the [Si\,{\sc vii}] emitting
region.  In contrasts, [Si\,{\sc vi}]
(not shown in the Figure~\ref{clr}) is limited to
the central $\sim$70\,pc. These values,
however, are larger than those derived by Oliva et al. (1994),
who estimated that the size of the CLR in Circinus was only $\sim$10\,pc.
Due to the better sensitivity of our data, we show that 
the coronal gas extends farther out from the centre than previously 
found. For
comparison, note that the H$_{2}$ lines in Figure~\ref{clr}  are 
detected in all regions where spectra were extracted.

In NGC\,1068, both [Si\,{\sc vi}] and [Si\,{\sc vii}] extends 
from the unresolved nucleus to 90\,pc south and up to 40\,pc\,north
(bottom and upper left panels of Figure~\ref{clr}).
These values, however, are significantly lower than that of 
$\sim$300\,pc to the south and 200\,pc to the north derived from 
 {\it NICMOS} [Si\,{\sc vi}] imaging of Thompson et al. (2001). Note,
however, that the Thomson et al's data may be misleading
because of contamination of [Si\,{\sc vi}] 1.963$\mu$m  
by H$_{2}$ 1.957$\mu$m, easily seen in Figure~\ref{clr}, 
where both lines are heavily blended. At 88\,pc from
the nucleus, [Si\,{\sc vi}] 1.963$\mu$m and H$_{2}$ 1.957$\mu$m 
have similar strengths and at distances larger than 120\,pc,
only the H$_{2}$ is detected. 

Overall, our data show that all CL are emitted in the 
inner 100\,pc of the active nucleus. When compared to low-ionization and
molecular lines, we found that the emitting region of the latter two 
set of lines extends to much larger distances (R$>$ 500\,pc). 
This result rules out the hypothesis of Ferguson et al. (1997) 
of a a low-density CLR extending up to 1\,kpc from the active 
nucleus but also rules out very compact CLR with upper limits
of a few parsecs. 

\section{Are pure photoionization driving the CL emission?} \label{mod}

We tested the possibility of the CL being emitted
from distances of up to several tens of parsecs from the 
central engine assuming pure photoionization. For this purpose,
we run models 
with the AANGABA code (Gruenwald \& Viegas 1992). The adopted spectral energy
distribution (SED) was that suggested by Fig.~7 of Oliva et al. (1999)
for Circinus. Two values of luminosities of the ionizing radiation were
adopted: $L_{\rm ion}=10^{43.5}$ erg s$^{-1}$ and
and $L_{\rm ion}=10^{44.5}$ erg s$^{-1}$, suggested by Ferguson 
et al. (1999) and Oliva et al. (1999), respectively. The number of
ionizing photons, $Q_{\rm H}$, provided by these the two SEDs
are 2.5$\times10^{53}$ s$^{-1}$and 2.5$\times10^{54}$ s$^{-1}$.
Density was varied from 10$^{2}$ cm$^{-3}$ to 10$^{6}$ cm$^{-3}$. 
Three values for the  ionization parameter $U$ were employed:
10$^{-3}$, 10$^{-2}$ and 10$^{-1}$.

\begin{figure}
\centering
\includegraphics[width=6cm]{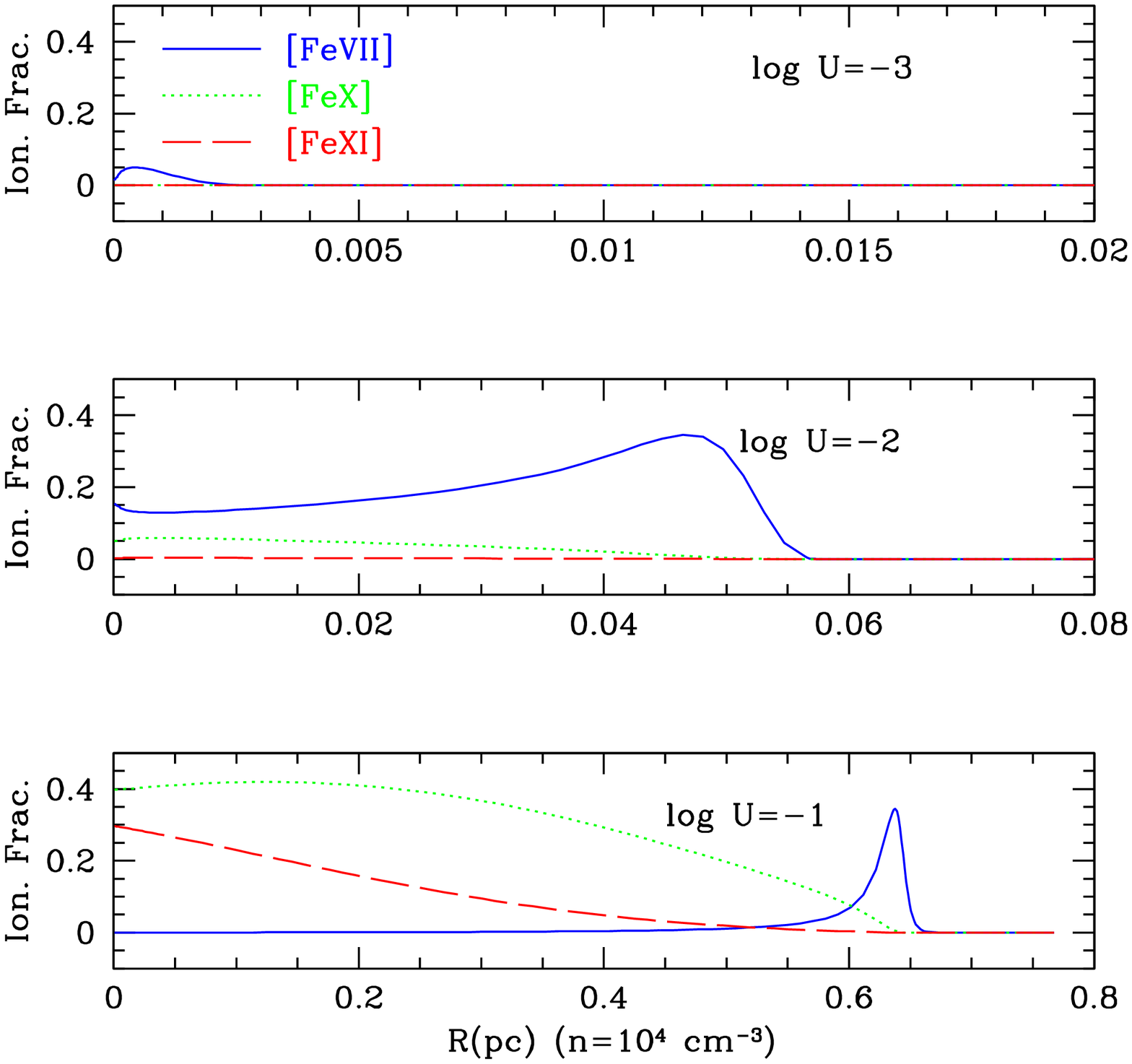}
\includegraphics[width=6cm]{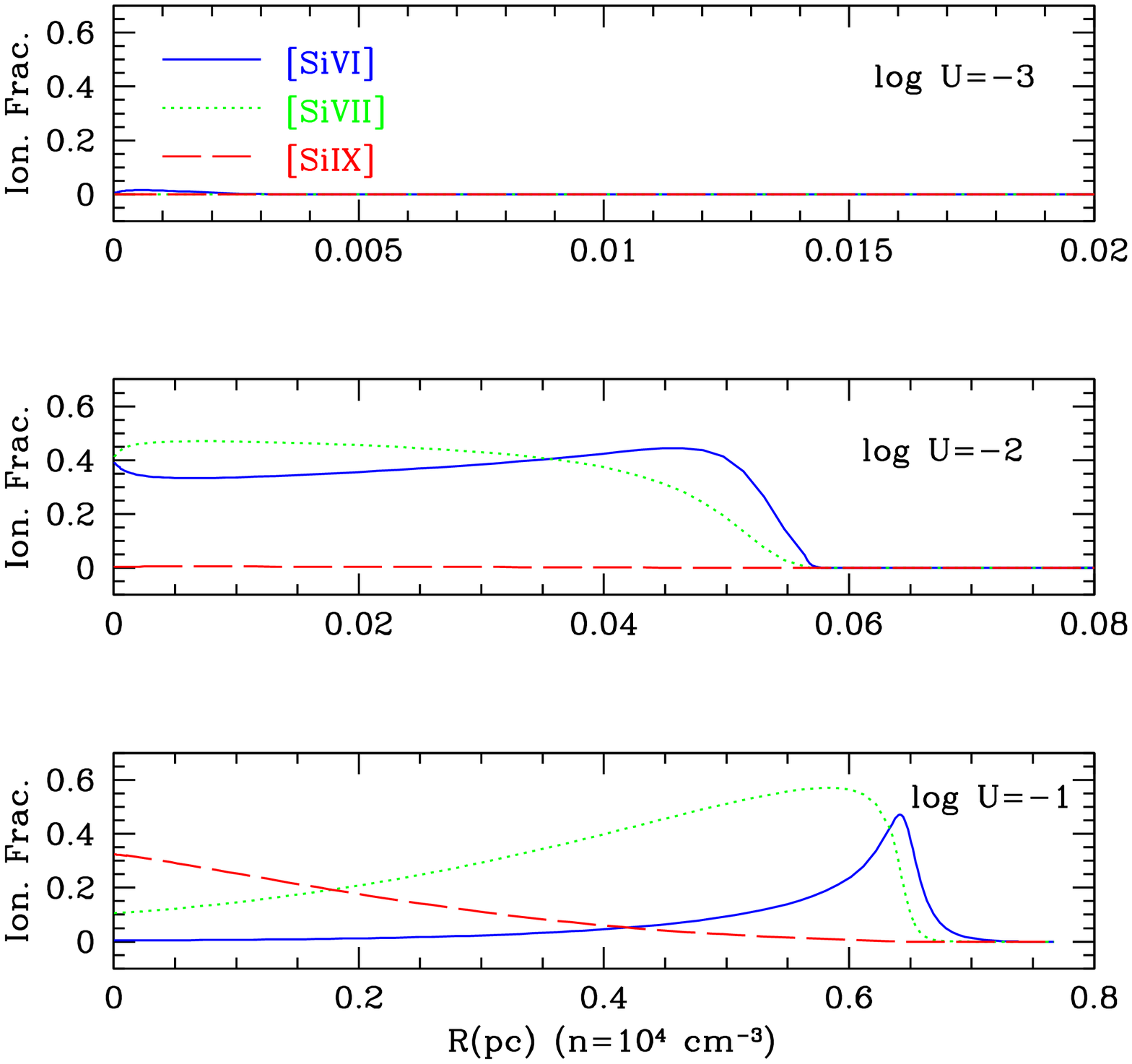}
 \caption{Ionic abundance for the Fe ions (left) and Si (right) 
versus the distance within the cloud for $L_{\rm ion} = 10^{43.5}$ 
erg s$^{-1}$. Typical cloud sizes are $\sim$1\,pc.} \label{ions}
\end{figure}

Figure~\ref{ions} shows the distribution of the ionic fraction 
of iron and silicon versus the size of the 
emitting cloud for the three values of $U$ employed. Note that 
R, equivelent to the cloud size, is measured from the side of 
the cloud facing the ionizing radiation 
to the outer edge. The value of density that best reproduces
the observations is $n_{\rm H}$=10$^{4}$ cm$^{-3}$,
in accord to that of 0.5$\times$10$^{4}$ cm$^{-3}$
determined for the CLR from the [Ne\,{\sc v}] 14.3$\mu$m / 24.3$\mu$m
line ratio with {\it ISO} (Moorwood et al. 1996).

The results show that an $U \geq 0.10$ is necessary in order
to have clouds with an ionization structure similar to
that observed $-$i.e., simultaneous presence of 
[Si\,{\sc vi}], [Si\,{\sc vii}], and [Fe\,{\sc vii}] through
[Fe\,{\sc xi}]. In addition, for the parameters adopted, the
emitting clouds must located within 8 to 30 pc from the
central engine. Coronal
emission located farther out from the source would requiere
$L_{\rm ion} > 10^{44}$ erg\,s$^{-1}$.  More detailed modeling 
of our data can be found in 
Rodr\'{\i}guez-Ardila et al. (2004, in preparation).

\section{Kinematics of the coronal gas} \label{kin}

\begin{figure}
\centering
\includegraphics[width=5.5cm]{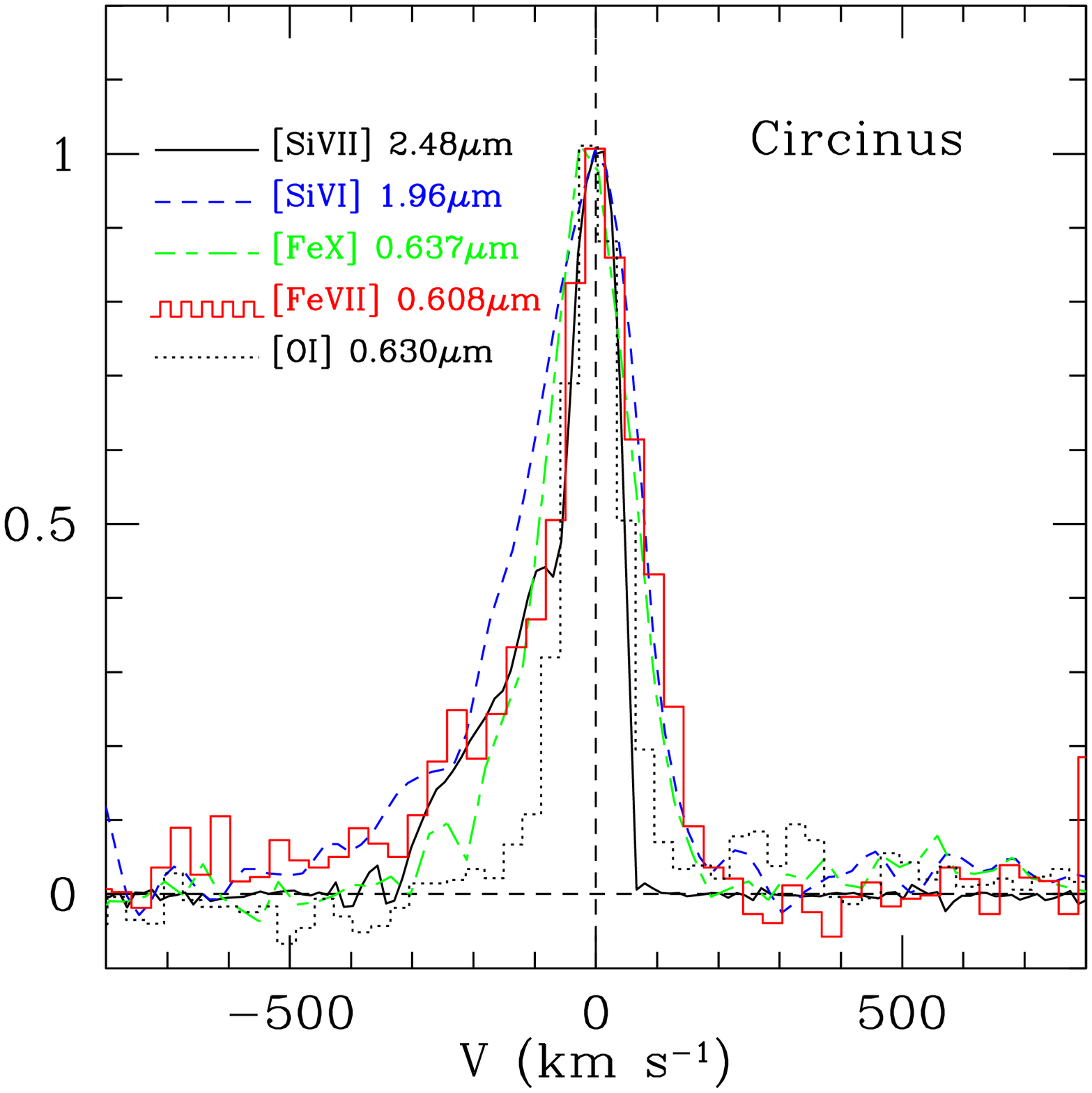}
\includegraphics[width=5.5cm]{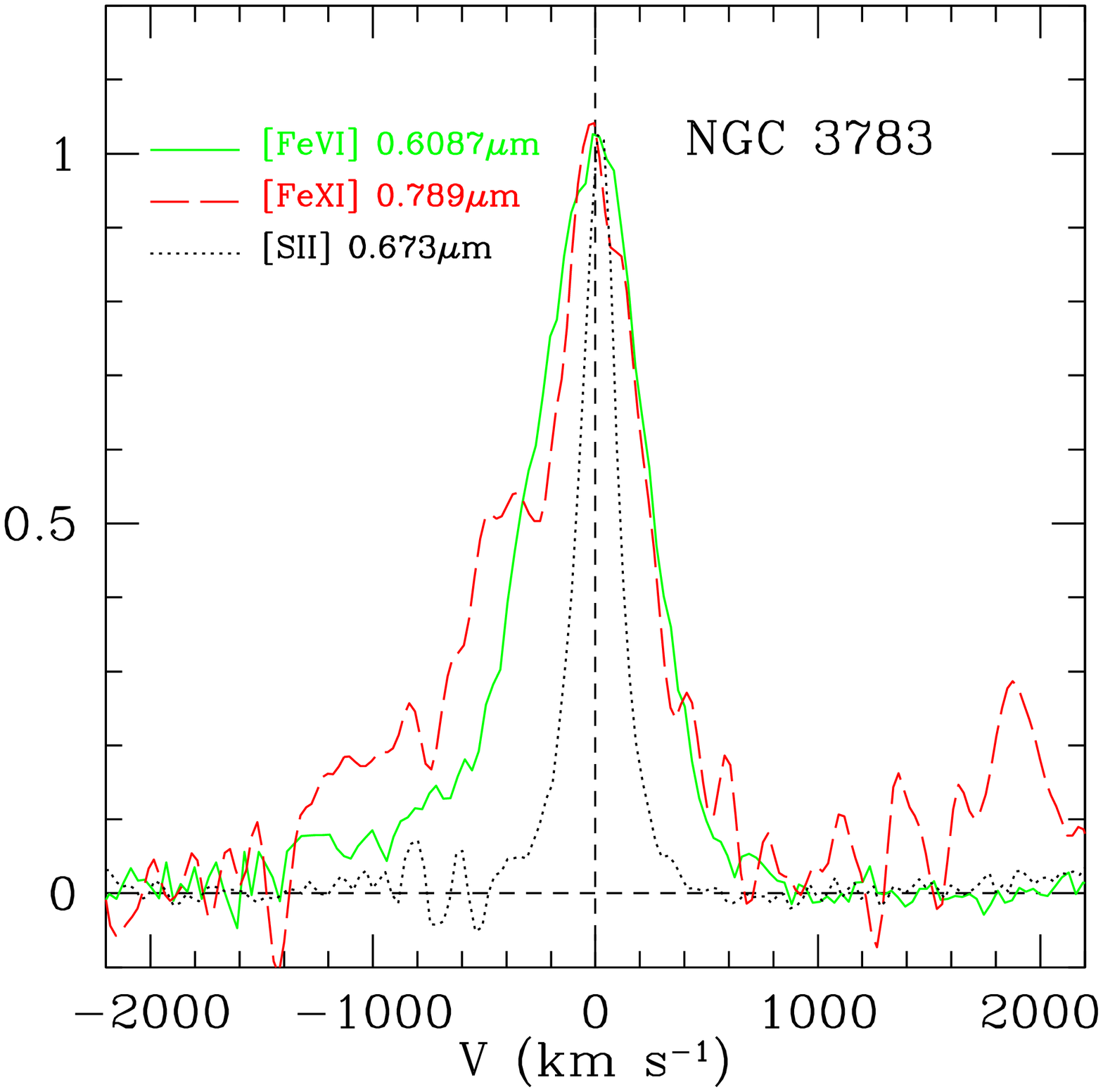}
  \caption{Comparison of low ionization and coronal emission line profiles,
in velocity space, for Circinus(left) and NGC\,3786 (right)}\label{clr:kin}
\end{figure}

The kinematics of the coronal gas was studied by comparing 
the emission line profiles of low and high
excitation lines. Figure~\ref{clr:kin} show the results obtained for
Circinus and NGC\,3783. 

As is easily seen, the coronal lines in both objects are highly 
asymmetric towards the blue and broader than low-ionization
lines. Previous works on Circinus (Oliva et al.
1994, 1999) had failed to detect any significant variation
between the width and the IP of the forbidden lines. In fact,
Moorwood et al. (1996) had reported narrower widths in 
high excitation lines than in lower excitation lines. The VLT/ISAAC
and NTT/EMMI spectra of Figure~\ref{clr:kin} remove this
ambiguity and shows the presence of coronal gas with velocities
close to 500 km\,s$^{-1}$ in Circinus.  Blue asymmetric profiles
are also observed in NGC\,3783. Note, however, that the line
profile of [Fe\,{\sc xi}] implies radial velocities of up to 
1500 km\,s$^{-1}$, suggesting as in Circinus, that part of the gas 
must arise in outflows very close to the central engine.
Of the six objects studied, only NGC\,1068 shows no difference
between the FWHM of low and high ionization lines.  

\section{Summary} \label{con}

EMMI/NTT and ISAAC/VLT spectra are used to determine
the size of the CLR in a sample of well-known Seyfert 1
galaxies by means of the simultaneous observation of [Fe\,{\sc vii}],
[Fe\,{\sc x}], [Fe\,{\sc xi}], [Si\,{\sc vi}] and
[Si\,{\sc vii}] lines. Our data show that all coronal 
lines are emitted from the unresolved nucleus to distances
of up to 80 pc.  We showed
that the coronal emission region can arise from
gas photoionized from radiation of the central engine if
$U\geq$0.1, $n_{\rm e} \sim$10$^{4}$ cm$^{-3}$, 
solar abundances, cloud sizes of $\sim$1 pc and 
distances from the central source of up to 30 pc. 
The analysis of the coronal emission line profiles shows
that they are highly asymmetric towards the blue, even in
Circinus, where previous observations had measured
widths of the order of 100 km\,s$^{-1}$. For this object,
we found coronal gas with velocities of up to 
400 km\,s$^{-1}$, suggesting that shocks must probably
contribute to the observed emission.

\end{document}